# Impact of a doping-induced space-charge region on the collection of photo-generated charge carriers in thin-film solar cells based on low-mobility semiconductors

Oskar J. Sandberg,[1] Staffan Dahlström,[2] Mathias Nyman,[2] Sebastian Wilken,[2,3] Dorothea Scheunemann,[2,3] and Ronald Österbacka[2]

[1]Department of Physics, Swansea University, Singleton Park, Swansea SA2 8PP Wales, United Kingdom

[2]Physics, Center for Functional Materials and Faculty of Science and Engineering, Åbo Akademi University, Porthaninkatu 3, 20500 Turku, Finland

[3]Institute of Physics, Energy and Semiconductor Research Laboratory, Carl von Ossietzky University of Oldenburg, 26111 Oldenburg, Germany

Unintentional doping of the active layer is a source for lowered device performance in organic solar cells. The effect of doping is to induce a space-charge region within the active layer, generally resulting in increased recombination losses. In this work, the impact of a doping-induced space-charge region on the current-voltage characteristics of low-mobility solar cell devices has been clarified by means of analytical derivations and numerical device simulations. It is found that, in case of a doped active layer, the collection efficiency of photo-generated charge carriers is independent of the light intensity and exhibits a distinct voltage dependence, resulting in an apparent electric-field dependence of the photocurrent. Furthermore, an analytical expression describing the behavior of the photocurrent is derived. The validity of the analytical model is verified by numerical drift-diffusion simulations and demonstrated experimentally on solution-processed organic solar cells. Based on the theoretical results, conditions of how to overcome charge collection losses caused by doping are discussed. Furthermore, the presented analytical framework provides tools to distinguish between different mechanisms leading to voltage dependent photocurrents.

## I. INTRODUCTION

Thin-film solar cells based on organic or organic-inorganic hybrid semiconductor systems hold great potential for future energy production. The lab scale power conversion efficiency has recently exceeded the 15 % mark for organic bulk heterojunction (BHJ) solar cells, and the record efficiency for organic-inorganic halide perovskite solar cells is currently at 23.7 % [1]. The impressive improvement in device efficiencies is largely due to new, better performing materials being developed and synthesized, in addition to an increased understanding of the processes and mechanisms leading to efficiency losses. In fact, a large part of the field of next generation thin-film solar cells is focused around the design and synthesis of new material blend systems [2].

The generic thin-film solar cell device structure is composed of a photoactive intrinsic layer sandwiched between two charge extracting electrode (inter)layers, the anode and the cathode; in organic solar cells, the active layer is a blend of a polymer or small-molecule, mixed with fullerenes or other polymers or small molecules [3-5]. The net photocurrent density is given by the difference between the current of the device under illumination ($J$) and in the dark ($J_{\text{dark}}$),

$$J_{ph} = J - J_{\text{dark}} \qquad (1)$$

Ideally, the net photocurrent is constant with the applied voltage $V$ and given by the short-circuit current density: $|J_{ph}| = J_{SC}$, where $J_{SC}$ is the magnitude of the short-circuit current density, corresponding to the current induced by light at short-circuit [6]. However, the relation $|J_{ph}| = J_{SC}$ is commonly only valid for solar cells based on high-mobility, non-excitonic, semiconductors (such as crystalline Si) where both the generation and the extraction of photo-induced charge carriers is independent of the voltage $V$. In low-mobility systems such as organic BHJ solar cells, on the other hand, neither of these two conditions are valid in general.

Firstly, owing to the inherent excitonic nature of organic systems, the generation and recombination of separated charge carriers in these materials is generally taking place via intermediate charge transfer states [7, 8]. In the past, the dissociation of these states into separated charge carriers has been considered an electric-field-assisted process in accordance with the classical Onsager-Braun model [3, 9], manifested as a field-dependent generation rate of separated charge carriers [10]. Similar types of electric-field dependent charge generation mechanisms have consequently also been used to explain voltage-dependent photocurrents in BHJ solar cells [11-14]. However, this electric-field dependence is generally found to be weak in state-of the-art BHJ blends, where the charge transfer state dissociation is efficient [15-19]. Secondly, the collection efficiency of photo-generated charge carriers in low-mobility materials depends on the ratio between the charge carrier recombination lifetime and the carrier extraction time $t_{\text{extr}} \sim d/\mu|F|$ [20-23], where $d$ is the thickness of the active layer, $F$ is the internal electric field, and $\mu$ is the mobility. In organic bulk heterojunctions where the competition between charge extraction and charge recombination of photo-induced carriers is important [24-28], an electric-field dependent charge collection efficiency is therefore to be expected. This type of voltage-dependent charge collection efficiency is also strongly influenced by space-charge effects, for example caused by imbalanced mobilities [29-32], which give rise to highly inhomogeneous electric field distributions inside the active layer.

Another source for space-charge effects is (unintentional) doping of the active layer [33-35]. Unintentional *p*-type doping of the active layer has frequently been encountered in organic solar cells, especially in thicker devices, and has been generally attributed to the presence of oxygen and water inside the active layer, or impurities due to residues from synthesis [36-39]. However, also diffusion of molecules from the electrode contacts and/or the electrode interlayers has been observed to cause doping of the active layer [40-44]. The effect of doping is to create a depleted space-charge region (SCR) within the active layer, adjacent to one of the contacts (a Schottky junction), at moderate doping levels [45]. For a *p*-doped active layer, containing a uniform concentration of (ionized) dopants $N_p$, the space-charge region is formed adjacent to the cathode. The thickness $w_0$ of this space-charge region is given by [46]

$$w_0 = \sqrt{\frac{2\varepsilon\varepsilon_0}{qN_p}\left[V_0 - \frac{kT}{q} - V\right]} \qquad (2)$$

assuming $0 < w_0 < d$, where $V$ is the applied voltage and $V_0$ is the built-in potential across the depletion region; $q$ is the elementary charge, $T$ is the temperature, $k$ the Boltzmann constant, $\varepsilon_0$ is the vacuum permittivity and $\varepsilon$ is the relative permittivity of the active layer. In this case, the electric field is mainly concentrated to the depletion region, leaving the rest of the active layer essentially charge-neutral and free of electric field. For solar cells with low mobilities, this inevitably leads to an inefficient collection of photo-induced carriers within the (quasi-)neutral region, where the charge extraction is driven by diffusion [34, 47-49].

Although the effect of doping has been recognized to result in a voltage dependent charge collection in organic solar cells, the resulting apparent electric field dependence of the photocurrent is often overlooked when analyzing and interpreting experimental photocurrents. When characterizing new materials, it is of particular importance to distinguish between different mechanisms leading to voltage dependent photocurrents, since in some cases the underlying reason is intrinsic to the material (field-dependent generation, poor extraction due to morphology, etc.) and in some cases extrinsic (doping caused by impurities, degradation, diffusion of dopants from contacts, etc.). In this work, the effect of a doping-induced space-charge region on the charge collection in low-mobility solar cells, with optically thin active layers, is clarified. An analytical expression is derived, explaining the voltage dependent behavior of the photocurrent. The analytical framework is verified by numerical drift-diffusion simulations. The analytical model explains the voltage dependence of experimental photocurrents observed in (unintentionally) doped organic solar cells based on P3HT:PCBM.

## II. THEORETICAL BACKGROUND

The model device consists of an optically thin active layer sandwiched between two charge extracting electrode layers, the hole-extracting anode (at $x = 0$) and the electron-extracting cathode (at $x = d$). The equations describing the electrical transport under steady-state conditions are [46, 50]

$$-\frac{1}{q}\frac{dJ_n}{dx} = G_L - R \tag{3}$$

$$\frac{1}{q}\frac{dJ_p}{dx} = G_L - R \tag{4}$$

with $G_L$ being the photo-induced generation rate of free carriers and $R$ the recombination rate, whereas the current densities for electrons and holes are assumed to be given by the drift-diffusion (or Nernst-Planck) equations:

$$J_n(x) = q\mu_n n F + qD_n \frac{dn}{dx} \tag{5}$$

$$J_p(x) = q\mu_p p F - qD_p \frac{dp}{dx} \tag{6}$$

respectively. Here, $n$ is the electron density and $p$ is the hole density, $\mu_n$ is the electron mobility, $\mu_p$ is the hole mobility, whereas $D_n$ and $D_p$ are the electron and hole diffusion coefficients, respectively. Furthermore, we assume that the diffusion coefficients are related to the mobilities via the classical Einstein relation $D_{n(p)} = \mu_{n(p)} kT/q$ [51].

The electric field $F$ within the active layer is determined via the Poisson equation,

$$\frac{dF}{dx} = \frac{\rho_{sc}}{\varepsilon\varepsilon_0} \tag{7}$$

where $\rho_{sc}$ is the net space-charge density in the organic semiconductor layer. In an active layer where *p*-doping is present (neglecting *n*-type doping), the space-charge density reads

$$\rho_{sc} = q[p - n - N_p] \tag{8}$$

where $N_p$ is the density of ionized (negatively charged) p-dopants. The p-dopants are commonly constituted by acceptor-like (i.e. negatively-charged when occupied by an electron and neutral when empty) impurity atoms or molecules [46, 52]. When the frontier energy levels of the impurity (such as the LUMO level of an impurity molecule) is close to the valence level (HOMO) of the organic semiconductor, an electron can be accepted directly from the HOMO level of the organic semiconductor, giving rise to p-type doping [52-54]. An analogous situation applies in case of n-type doping.

The total steady-state current density $J = J_n(x) + J_p(x)$ can be expressed as

$$J = J_n(0) + J_p(d) - q \int_0^d [G_L - R] \, dx \tag{9}$$

where the net bulk recombination rate of charge carriers generally takes the form [46, 55]

$$R = \beta(np - n_i^2) \tag{10}$$

where $\beta$ is the bimolecular recombination coefficient and $n_i^2 = N_c N_v \exp(-E_g/kT)$, where $E_g$ is the electrical bandgap of the separated charge carriers. The second term in Eq. (10) corresponds to the thermal generation rate of carriers in the dark,

$$G_{th} = \beta n_i^2 \tag{11}$$

Here, the effect of generation and recombination via charge-transfer states is assumed to be implicitly incorporated into the recombination coefficient $\beta$ [9, 56, 57].

To solve the set of coupled differential equations, a numerical 1D drift-diffusion model is used [58-60]. The thickness of the active layer is assumed to be 150 nm, with a dielectric constant of $\varepsilon = 3$, an electrical bandgap of $E_g = 1.2$ eV, and effective density of transport states given by $N_c = N_v = 10^{20}$ cm$^{-3}$ for both electrons and holes. The charge carrier generation rate is assumed to be uniform and *independent* of the electric field, the rate $G_L = 6.24 \times 10^{21}$ cm$^{-3}$/s corresponding to 1 sun. The extraction of holes and electrons at the anode and cathode contacts, respectively, is assumed ideal with an electron injection barrier $\varphi_n = 0.2$ eV at the cathode, whereas the Fermi level of the hole contact is assumed to be pinned by the doping level at the anode ($p(0) = N_p$). The corresponding built-in voltage is given by $qV_0 = E_g - \varphi_n - kT \ln(N_v/N_p)$. Finally, *unless otherwise stated*, the following parameters will be assumed in the simulations: a fixed recombination coefficient of $\beta = 1.2 \times 10^{-11}$ cm³/s for bulk recombination, selective contacts with $J_p(d) = J_n(0) = 0$, charge carrier mobilities of $\mu_n = \mu_p = 10^{-3}$ cm²/Vs, and a hole doping concentration of $N_p = 10^{17}$ cm$^{-3}$.

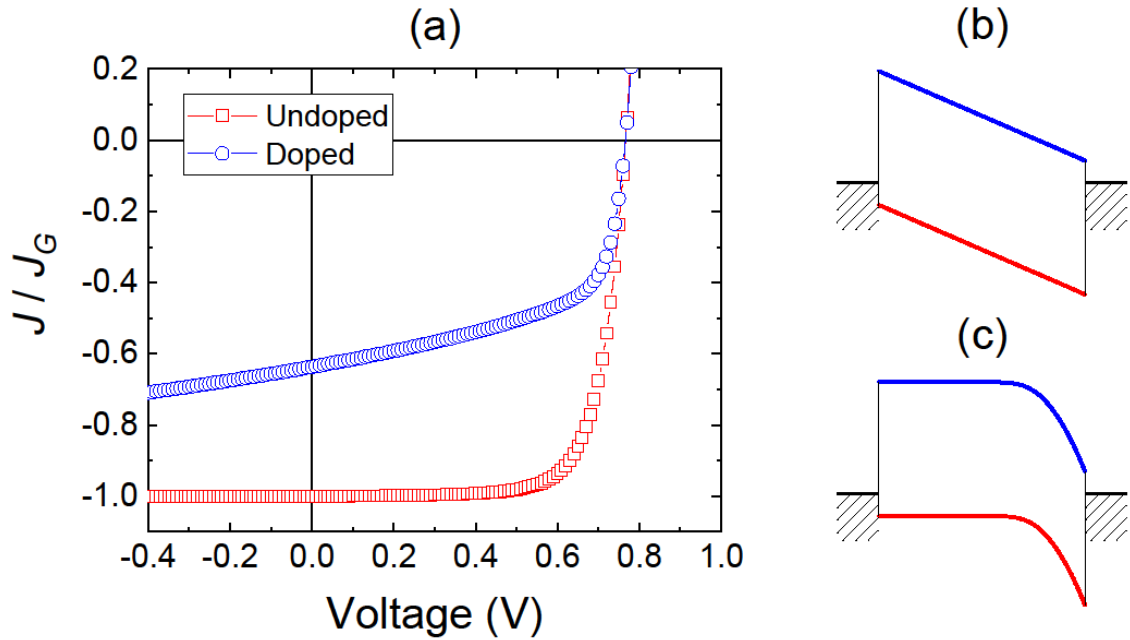

**Fig. 1: In (a), the current-voltage characteristics for a device under light is simulated for the case with an undoped (red squares) and a doped active layer (blue circles). In (b) and (c), the corresponding energy level diagrams at thermal equilibrium are depicted for the case of the undoped (intrinsic semiconductor) and the *p*-doped active layer, respectively. A doping concentration of $N_p = 10^{17}$ cm$^{-3}$ is assumed for the doped case.**

## III. RESULTS

The simulated *J-V* curves of a solar cell device with an undoped and a *p*-doped active layer are shown in Fig 1. The current densities have been normalized to the average photo-generation current density $J_G \equiv qG_L d$, corresponding to the magnitude of the photocurrent density obtained under complete charge extraction. The corresponding energy level diagrams are shown in Fig 1(b) and (c). The effect of doping is to reduce the magnitude of the photocurrent. Whereas most of the photo-generated charge carriers are collected near short-circuit conditions in the undoped device, in the case of the *p*-doped device a strongly voltage-dependent charge collection is obtained. This is a direct consequence of the inefficient charge collection of carriers generated in the neutral region, which is present in the doped device; cf. Fig. 1(b) and (c). To understand the physics behind the current-voltage behavior obtained in case of doping in greater detail, in the following, we first investigate the situation from an analytical point of view.

### *1. The analytical model*

We consider a *p*-doped active layer with $N_p$ high enough for $0 < w_0 < d$ to be valid under operating conditions. A schematic energy level diagram is depicted in Fig. 2(a). Under illumination, the carrier extraction within the space-charge region is expected to be efficient because of the large electric field being concentrated to this region. It should be noted, however, that the electric field strength inside the space-charge region increases linearly with $x$, with $F = 0$ at $x = d - w_0$. Since the electric field in the beginning of the SCR is initially small, the region of efficient charge collection (where the electric field is strong enough for most electrons to be extracted), effectively assumed to be of thickness $w$, within the space-charge region, is in general expected to be smaller than $w_0$.

Taking the bulk recombination rate of photo-generated electrons to be negligible within the distance $w$ from the cathode, inside the space-charge region, it follows from Eq. (3) that $J_n(d) = J_n(0) - qG_L w -$

$\int_0^{d-w}[G_L - R]\,dx$. In the neutral region, the charge collection is driven by diffusion, with the SCR virtually acting as a sink for electrons: $n(d - w) \approx 0$. Then, noting that the recombination rate can be approximated as $R \approx \beta N_p n$, the electron continuity equation for $0 < x < d - w$ reads

$$-\frac{1}{q}\frac{dJ_n}{dx} = -\frac{\mu_n kT}{q}\frac{d^2 n}{dx^2} = G_L - \frac{n}{\tau} \tag{12}$$

where

$$\tau = \frac{1}{\beta N_p} \tag{13}$$

is the recombination lifetime for electrons *in the neutral region*. From the solution to Eq. (12), and assuming surface recombination of electrons at the anode to be negligible $J_n(0) = 0$, the net photocurrent is readily obtained as

$$J_{ph} = -qG_L\left[w + L_n \tanh\left(\frac{d-w}{L_n}\right)\right] \tag{14}$$

for $w < d$, where

$$L_n = \sqrt{\frac{\mu_n \tau kT}{q}} \tag{15}$$

is the characteristic diffusion length for electrons within the neutral region. Based on these considerations (neglecting leakage currents induced by parasitic shunts), the total current density under illumination can be approximated as

$$J = J_{ph} + J_0\left[\exp\left(\frac{qV}{kT}\right) - 1\right] \tag{16}$$

for $V < V_{bi} - kT/q$, where $J_0 = qG_{th}[w + L_n \tanh([d - w]/L_n)]$ after noting that the dark saturation current density $J_0$ is equal to $|J_{ph}|$ when $G_L = G_{th}$.

An explicit expression for the extraction length $w$ can be obtained based on the following approximations:

i) the diffusion-induced hole density tailing into the SCR is given by $p(x) \approx N_p \exp(-[x - d + w_0]^2/2L_D^2)$ for $x \geq d - w_0$, where

$$L_D = \sqrt{\frac{\varepsilon\varepsilon_0 kT}{q^2 N_p}} \qquad (17)$$

is the associated Debye screening length; ii) for $x \geq d - w$, the recombination is assumed negligible and $dJ_n/dx = -qG_L$, and iii) for $x \leq d - w$, we have $dJ_{n,di}/dx = -q(G_L - R)$. At $x = d - w$, we thus require that $-dJ_{n,dr}/dx = q\beta np$, where $dJ_{n,dr}/dx \approx qn\mu_n\, dF/dx = (q^2 n\mu_n/\varepsilon\varepsilon_0)[p - N_p]$, assuming $n \ll N_p$ to be constant within in this region. Here, $J_n = J_{n,dr} + J_{n,di}$ with $J_{n,dr}$ and $J_{n,di}$ being the drift and diffusion components of the electron current, respectively. Subsequently, the effective extraction length is then obtained as

$$w = w_0 - \Delta w_R \qquad (18)$$

where

$$\Delta w_R = L_D\sqrt{2 \ln\left(1 + \frac{L_D^2}{L_n^2}\right)} \qquad (19)$$

is the correction term accounting for the recombination near the boundary between the effectively recombination-free SCR and the neutral region. Note that only in the limit of negligible recombination (i.e. $L_n \gg L_D$), $w \to w_0$, as expected.

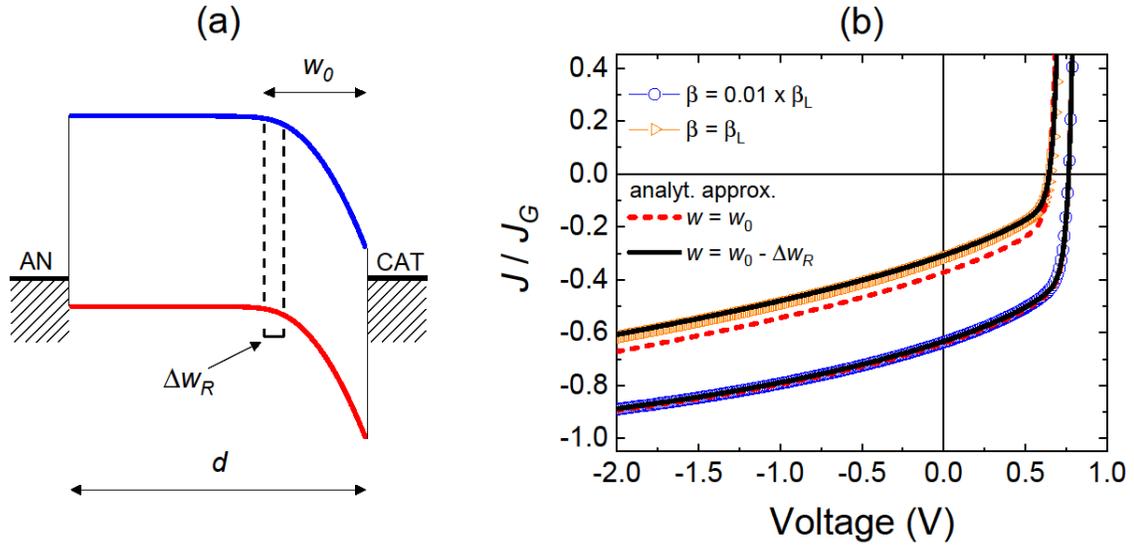

**Fig. 2:** In (a) the energy level diagram at $V = 0$ of the *p*-doped device under consideration, having a depletion region thickness of $w_0$, is shown. In (b), the simulated *J-V* curve of the *p*-doped solar cell under illumination from Fig. 1(a) is plotted for a larger voltage interval, and indicated by the blue circles. For comparison, also the case with a 100 times larger recombination coefficient ($\beta = \beta_L \equiv 1.2 \times 10^{-9}$ cm³/s) has been included, as indicated by the orange triangles. The analytical expressions Eq. (16), with $w$ given by $w_0$ [Eq. (2)] and $w = w_0 - \Delta w_R$ [Eq. (18)] are indicated by the corresponding red dashed lines and black solid lines, respectively.

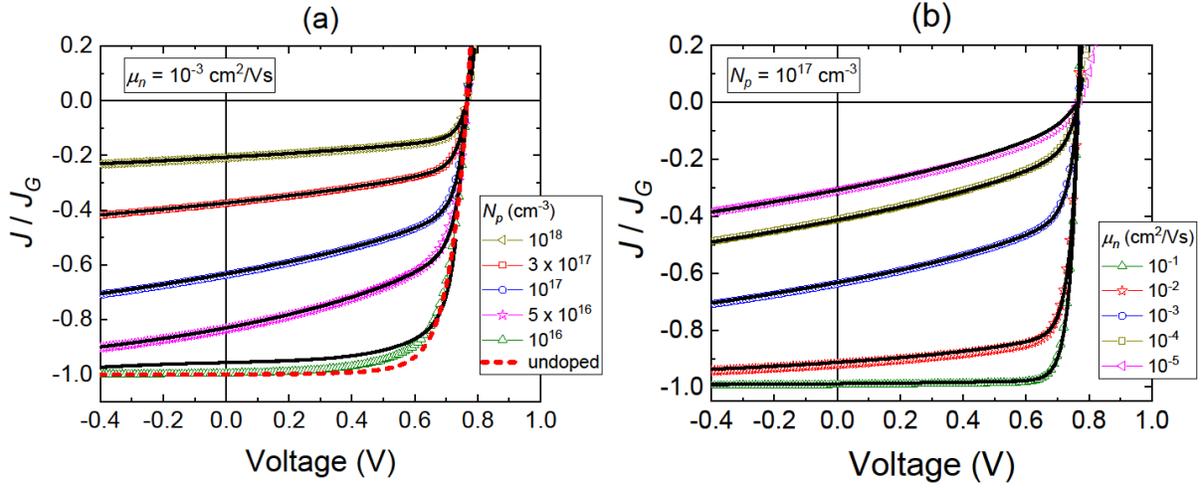

**Fig: 3.** Simulated current-voltage characteristics under light are shown for solar cell devices with doped active layer in (a) for varying doping concentrations at a fixed electron mobility of $\mu_n = 10^{-3}$ cm²/Vs, and (b) with varying electron mobility $\mu_n$, but a fix doping concentration of $N_p = 10^{17}$ cm⁻³. The solid black lines correspond to the analytical expression Eq. (16), with $w = w_0 - \Delta w_R$. The recombination coefficient and hole mobility is fixed at $\beta = 1.2 \times 10^{-11}$ cm³/s and $\mu_p = 10^{-3}$ cm²/Vs, respectively.

*2. The influence of recombination in the active layer*

The above theoretical analysis predicts the photocurrent to be given by Eq. (14) and the total current density to be given by Eq. (16). Furthermore, $w$ is found to be given by $w = w_0 - \Delta w_R$ in accordance with Eq. (18). In Fig. 2(b), the analytical prediction, depicted by the solid lines, is compared to the simulated *J-V* curves (circles) of the *p*-doped device (from Fig. 1(c)). For comparison, also the case with a 100 times larger recombination coefficient is included. Indeed, an excellent agreement is obtained. It can also be seen that assuming $w = w_0$ in Eq. (16) (red dashed line in Fig. 2(b)), and thus not correcting for the recombination in the depletion zone, will lead to a deviation between Eq. (16) and the simulations at larger recombination rates. In accordance with the analytical model, for a *p*-doped device, only electrons photo-generated within the distance $w + L_n \tanh([d - w]/L_n)$ from the cathode contact are collected. This distance changes with the applied voltage (via $w$), giving rise to a $J_{ph} \propto -\sqrt{(V_{bi} - kT/q - V)}$ behavior, thus explaining the observed apparent electric field dependence of the photocurrent as seen for the $p$-doped devices in Fig. 1 and Fig. 2. We note that this type of voltage-dependent photocurrents might easily be mistaken for an electric field dependent generation rate. Note that a *field-independent* generation rate has been assumed in all of the simulations.

In Fig. 3, the *J-V* curves of a *p*-doped device under illumination is simulated for varying electron mobilities and doping concentrations in the active layer. In Fig. 4, on the other hand, the charge collection efficiency $J_{SC}/J_G$ for a *p*-doped device under short-circuit conditions is simulated for varying light intensity and electron mobility. Upon comparing the analytical model Eq. (14) and Eq. (16) (where $w = w_0 - \Delta w_R$) with the simulations in Figures 3 and 4, a good overall agreement is obtained. In particular, at large doping concentrations, high mobilities, and low light intensities, the agreement between the simulations and the analytical model is excellent.

At large doping concentrations, the recombination lifetime for electrons inside the neutral region is low giving rise to very short electron diffusion lengths $L_n$. However, simultaneously, also the space-charge region is thinner, resulting in a much stronger electric field in the SCR. In accordance with Eq. (14), the photocurrent can under these conditions ($L_n \ll d - w$) be approximated by $J_{ph} = -qG_L[w_0 - \Delta w_R + L_n]$. Also, since the voltage dependence of $w_0$ scales inversely with $N_p$, a less pronounced electric field dependence is obtained at high doping concentrations. Conversely, at high mobilities when the electron diffusion length $L_n$ is large, conditions when the charge transport dominates over the bulk recombination are established, and the current saturates to $J_{ph} = -J_G$. Note that a similar saturation is also expected to eventually occur under large reverse bias, when $w$ becomes comparable to $d$. It should be noted that a deviation from the analytical prediction Eq. (16) is expected to occur at low enough doping concentrations when the active layer is fully depleted, and the device effectively becomes undoped.

In accordance with Eq. (14) and the simulations in Fig. 4(a), a linear dependence between the photo-generation rate and the photocurrent, $J_{ph} \propto G_L$, is obtained in case of a *p*-doped active layer at not too high light intensities ($n \ll N_p$), manifested as a light intensity *independent* charge collection efficiency. This can be traced back to the fact that the recombination rate of photo-generated electrons inside the neutral region is effectively first-order, predominantly taking place between photo-generated electrons and doping-induced holes (of fixed density), in accordance with Eq. (12). At higher intensities, however, a deviation from Eq. (14) will eventually take place as i) bimolecular losses in the SCR becomes important, and/or ii) the photo-induced carrier density within the neutral region becomes larger than the background doping concentration of holes. On the other hand, as seen from Fig. 4(b), the agreement with Eq. (14) is also excellent over a wide range of electron mobilities, provided that the contacts are selective. For comparison, also the analytical approximation with $w = w_0$ has been included in Fig. 4(b), showing a deviation at smaller mobilities. Hence, correcting for the recombination in the depletion region, i.e. $w = w_0 - \Delta w_R$, becomes important at low mobilities.

It should be emphasized that, in the above analysis, perfectly selective contacts have been assumed [61]. In case of contacts composed of metallic or highly conductive electrode interlayers [62], however, the selective extraction of only holes at the anode can no longer be guaranteed, and surface recombination of minority carriers at the electrodes starts to play a role [57, 63-68]. In case of non-selective contacts, the surface recombination of electrons at the anode becomes significant at high mobilities ($\mu_n > 10^{-3}$ cm$^2$/Vs) when the electron diffusion length $L_n > (d - w)/2$, as shown in the Supplemental Material [69].

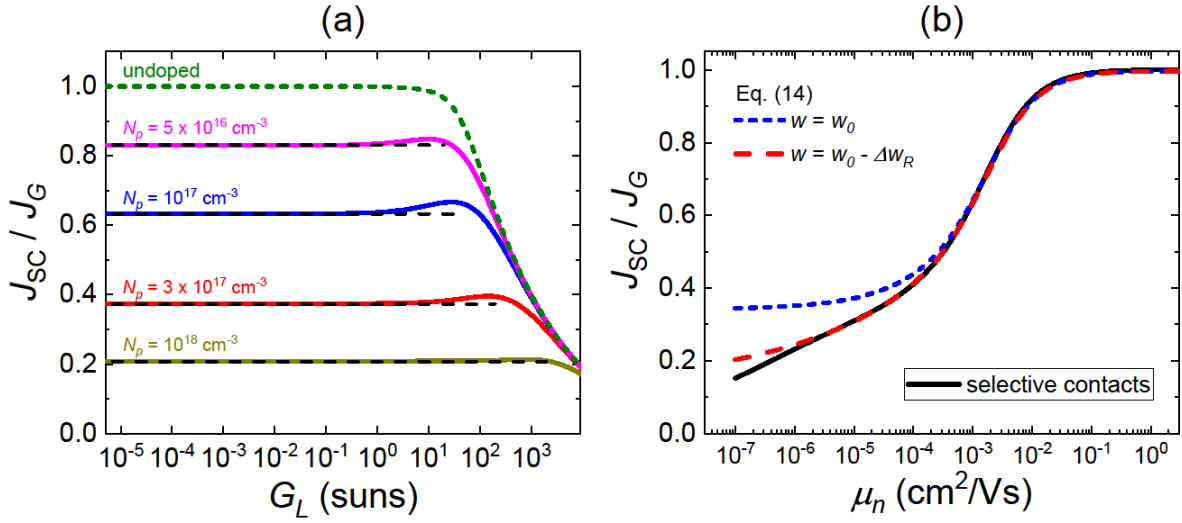

**Fig. 4:** The collection efficiency under short-circuit conditions $J_{SC}/J_G$ is simulated as (a) function of the generation rate at different doping concentrations, and (b) as function of the electron mobility $\mu_n$ at 1 sun light intensity and doping concentration $N_p = 10^{17}$ cm$^{-3}$. The analytical expression Eq. (14) is indicated by the corresponding black dashed lines in (a). The recombination coefficient and the hole mobility is fixed at $\beta = 1.2 \times 10^{-11}$ cm³/s and $\mu_p = 10^{-3}$ cm²/Vs, respectively.

*3. Influence of the charge transport of majority carriers in the neutral region*

In the above considerations, the conductivity of majority carriers within the neutral zone, i.e. of the holes, has been assumed to be large enough in order to not limit the current. To clarify the role of the hole conductivity, the impact of the hole mobility on the *J-V* curves in case of a *p*-doped active layer under illumination is simulated in Fig 5(a). Surprisingly, upon reducing the hole mobility in Fig. 5(a), an overall improved device performance can be initially obtained in this case. Following this initial increase, however, the charge collection efficiency eventually starts to decrease with decreasing hole mobilities.

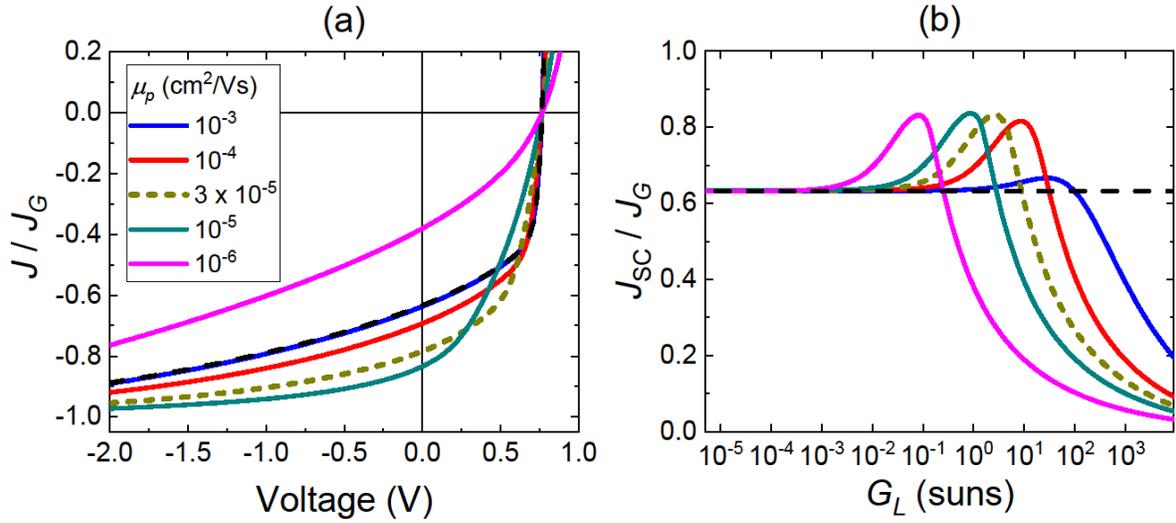

**Fig. 5:** (a) Simulated current-voltage characteristics at 1 sun light intensity of a *p*-doped device for varying hole mobilities $\mu_p$. The electron mobility is fixed at $\mu_n = 10^{-3}$ cm²/Vs and $N_p = 10^{17}$ cm⁻³. The analytical expression Eq. (16) is indicated by the black dashed line. In (b), the corresponding charge collection efficiency $J_{SC}/J_G$ under short-circuit conditions as a function of the light intensity $G_L$ is shown for the different $\mu_p$ from (a).

The reason for the initial enhancement of the charge collection efficiency with decreasing majority carrier mobility can be rationalized as follows. In the neutral region, the hole current is approximately given by $J_{ph} \approx J_p = q\mu_p N_p F$, assuming diffusion to be negligible *for holes* and neglecting surface recombination of electrons ($J = J_p(d)$). Subsequently, the magnitude of the principal electric field inside the neutral region can be approximated as

$$|F| \approx \frac{|J_{ph}|}{q\mu_p N_p} \qquad (20)$$

When the hole conductivity, i.e. the product $\mu_p N_p$, within the neutral zone is large enough for $|F| \ll kT/qd$, the electric field within the neutral region is negligible. Under such conditions, the electron transport inside the neutral zone is governed by diffusion and Eq. (14) describes the photocurrent well. However, at small hole mobilities (and/or low doping concentrations), the magnitude of the electric field inside the neutral region, in accordance with Eq. (20), will eventually become large enough for $|F| \gg kT/qd$. When this occurs, the charge transport of electrons within the neutral region becomes instead dominated by drift. Because of this field-assisted charge extraction, the average collection time for electrons within in the neutral region is shorter thus resulting in an increased charge collection efficiency.

In Fig. 5(b), the short-circuit current density is shown as a function of the generation rate at the different hole mobilities from Fig. 5(a). The corresponding magnitude of the electric field $|F(0)|$ within the neutral region (at the anode contact) is simulated in Fig. 6. As expected, the charge collection of photo-induced carriers within the neutral region is dominated by drift when

$$|J_{ph}| > \frac{\mu_p N_p kT}{d} \qquad (21)$$

Hence, depending on the generation rate and the doping concentration, an increase in the photocurrent relative to Eq. (14) can be obtained by reducing the hole mobility in case of a *p*-doped active layer. In other words, when Eq. (21) applies, the photocurrent is no longer given by Eq. (14). However, at even smaller hole mobilities or high enough generation rates, eventually, the density of the sluggish holes become comparable to (and larger than) the background doping concentration $N_p$. This is manifested by a drastic increase in the bulk recombination resulting in reduced current levels and degraded charge collection in Figure 5.

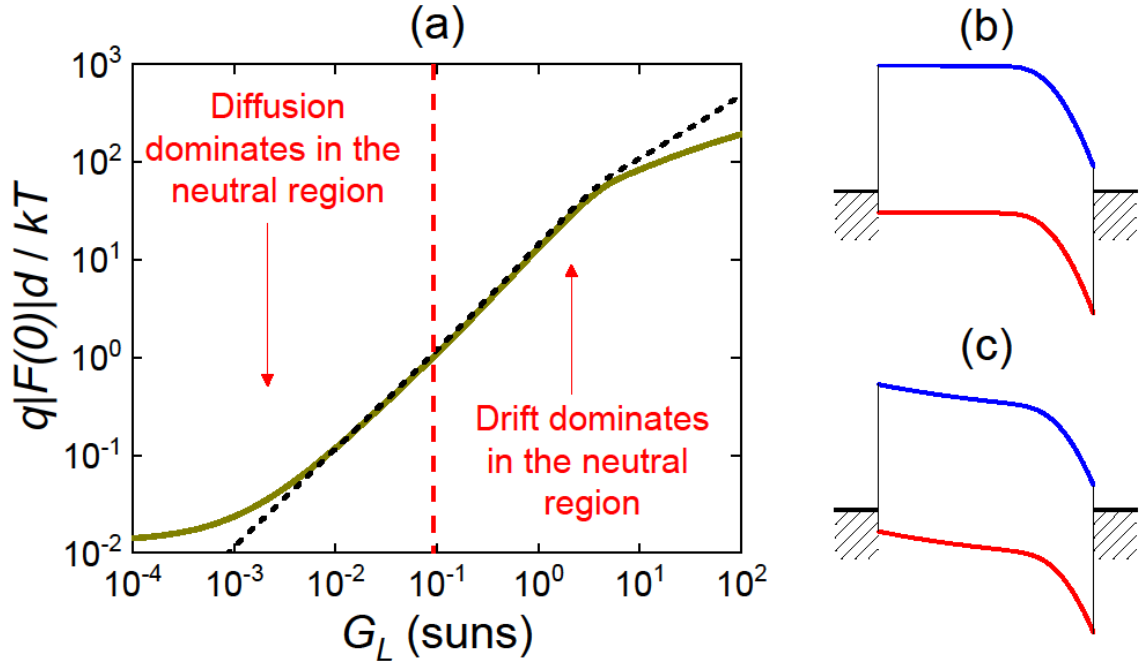

**Fig. 6:** (a) The magnitude of the electric field $|F(0)|$ in the neutral region (at the anode contact) under short-circuit conditions is simulated (solid line) as a function of $G_L$ for the case $\mu_p = 3 \times 10^{-5}$ cm²/Vs from Fig. 5. The electric field is normalized to $kT/qd$. The black dashed line depicts the case assuming $|F(0)|$ to be given by Eq. (20). The electron mobility is fixed at $\mu_n = 10^{-3}$ cm²/Vs and $N_p = 10^{17}$ cm⁻³. In (b) and (c), the corresponding energy level diagrams under short-circuit conditions are shown at 0.01 suns (diffusion dominates in the neutral region) and at 1 sun (drift dominates in the neutral region), respectively.

*4. Comparison with experiments*

In Fig. 7(a), experimental *J-V* curves of an organic solar cell device based on an active layer blend of poly(3-hexylthiophene-2,5-diyl) (P3HT) and [6,6]-phenyl C61 butyric acid methyl ester (PCBM) is shown for various light intensities. The device structure is ITO/MoO₃/P3HT:PCBM/LiF/Al, with the thickness of the organic active layer being 250 nm as measured by atomic force microscopy. The device fabrication and the *J-V* measurement setup are described in detail in [69]. The photocurrents show a pronounced voltage dependence under reverse bias. Given that the electric-field dependence of the charge carrier generation in this type of polymer:fullerene blend is known to be weak [13, 70], the observed voltage dependence of the photocurrent is attributed to inefficient charge collection. On the other hand, unintentional doping of P3HT:PCBM has been frequently seen in the past, in particular for rather thick devices [39, 41, 44]. By plotting the photocurrent at low intensities as a function of the square root of the applied voltage in reverse bias $\sqrt{-V}$, a linear dependence is seen at applied voltages much higher than $V_0 - kT/q$ (see Fig. S3 in the [69]). This type of linear dependence is expected in the case of a voltage-dependent photocurrent caused by doping [as per Eqs. (2) and (14)], and can thus be considered as an indication of a doped active layer.

In order to determine the doping concentration and the built-in voltage, capacitive charge extraction by linearly increasing voltage measurements are performed [71-73]. The details regarding the capacitive current measurements are given in [69]. The measured extraction current transients show features of a doped active layer, with the extracted depletion layer capacitance following a $C = \varepsilon\varepsilon_0/w_0$ behavior. Mott-Schottky analysis of the capacitive extraction current transients reveals a doping concentration of $N_p = 1.4 \times 10^{17}$ cm$^{-3}$ and a built-in potential of $V_0 = 0.77$ V, corresponding to a depletion layer thickness of $w_0 = 45$ nm under short-circuit conditions, in agreement with previous reports [41, 71]. Similar depletion region thicknesses have also been found in other systems using Mott-Schottky analysis of impedance spectra [36, 45, 48]. In ref 71 the origin of the doping was attributed to oxygen. However, in a more recent work on similar devices, it was shown that the doping within the active layer is caused by diffusion of MoO3 molecules into the active layer, acting as p-dopants inside the active layer [41]. We note that diffusion of MoO3 is most likely the predominant origin for the doping observed in this work and in Ref. [71] as well, given that a MoO3 interlayer was used in both cases.

The measured *J-V* curve at 0.06 suns is fitted using the analytical expression in Eq. (14) with $L_n$ and $G_L$ as fitting parameters. The fitted data is plotted in Fig. 7(b) and the analytical expression is able to describe the measured data very well. The measured dark current in reverse bias has been subtracted from the photocurrent in the plot. A low intensity of 0.06 suns is used for the fitting in order to ensure both $n \ll N_p$ and $|J_{ph}| < \mu_p N_p kT/d$ (see Eq. (21)), and thereby the validity of the analytical expression Eq. (14). The values $G_L = 1.6 \times 10^{20}$ cm$^{-3}$/s (at 0.06 suns) and $L_n = 80$ nm, used in the fit, are reasonable values for this material system [74, 75]. Assuming a hole mobility of $\mu_p = 3 \times 10^{-4}$ cm²/Vs [76], we obtain $\mu_p N_p kT/d \approx 7$ mA/cm². This is to be compared to the short-circuit current densities of $J_{SC} = 0.32$ mA/cm² and $J_{SC} = 5.8$ mA/cm² at 0.06 suns and 1 sun, respectively, confirming that Eq. (14) may be considered valid at 0.06 suns.

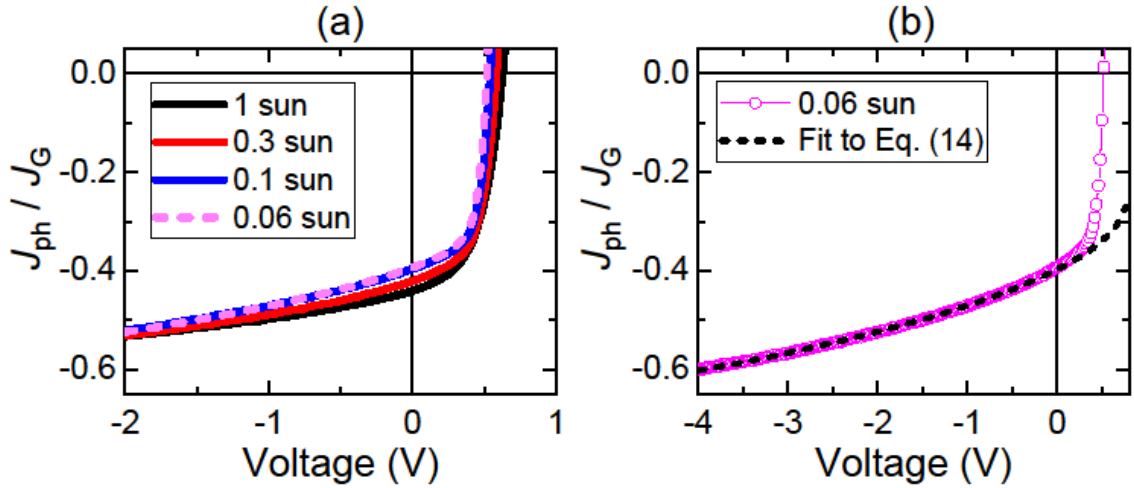

**Fig. 7: (a) Experimental *J-V* characteristics of a (unintentionally) doped P3HT:PCBM solar cell at four different light intensities. (b) *J-V* curve at 0.06 sun light intensity. The black dotted line corresponds to the analytical fit using Eq. (14). In both (a) and (b), the current densities have been normalized to their respective saturated photocurrent density $J_G$ as extracted from the analytical fit in (b).**

## IV. DISCUSSION

Based on the above theoretical findings, we see that the critical parameter governing the charge collection in *p*-doped organic solar cells is given by $(w + L_n)/d$. On average, only charge carriers photo-generated within the distance $w + L_n$ from the cathode is collected to the outer circuit. Subsequently, an efficient charge collection is in general only limited to the space charge region when the diffusion length $L_n \ll d$, corresponding to a large recombination rate inside the bulk. This also suggests that photocurrent is governed by the average charge carrier generation rate within the distance $w + L_n$ from the cathode in this case; for uniform carrier generation profiles, the photocurrent is thus independent of the active layer thickness (see Fig. S2).

However, if the mobility and/or recombination lifetime of the minority carriers within the neutral region is large enough so that $L_n > d - w_0$, then most of the charge carriers are extracted and charge collection losses caused by a doping-induced space charge region are small. In the linear intensity regime, this corresponds to doping concentrations of

$$N_p < \frac{\mu_n kT}{q\beta d^2}\left[1 - \frac{w_0}{d}\right]^{-2} \tag{22}$$

for $w_0 < d$, and simplifies as $N_p < \mu_n kT/q\beta d^2$ when $w_0 \ll d$. The effect of doping is thus less prominent in materials with low recombination coefficients and large minority carrier mobilities (see Fig. 3(b)). This restriction, however, can be relaxed by adjusting the mobility of the majority carriers (or rather their conductivity) in the bulk to such an extent that $|J_{SC}| \sim 10\,\mu_p N_p kT/d$. Hence, a slight reduction of the majority carrier mobility might in some cases be beneficial for the charge collection of minority carriers from the neutral region in doped active layers.

The doping-induced recombination losses effectively behave as a first order process. This makes it challenging to distinguish doping-induced losses from other first-order recombination losses such as geminate recombination. Our theoretical model, however, provides a tool of how to distinguish between a doping-induced apparent electric field dependence in the photocurrent from other field-dependent effects, such as an electric field dependent charge carrier generation rate. Provided that $|J_{ph}| < \mu_p N_p kT/d$ corresponding to low enough light intensities [the linear photocurrent regime in Fig. 5(b)], we expect

$$J_{ph} \approx -qG_L[w_0 + L_n] \tag{23}$$

for $L_n \ll d - w$.

Subsequently, if the field dependence of the photocurrent is dominated by doping-induced space-charge-limited carrier collection, a $J_{ph}$ versus $w_0$ plot should be linear. At reverse-bias voltages $|V| \gg V_0 - kT/q$, this translates into the linear dependence between $J_{ph}$ and $\sqrt{-V}$ (for $V < 0$) demonstrated in Fig. S3(a). Here, $w_0(V)$ can be readily obtained from Mott-Schottky analysis of the capacitance-voltage measurements (at low frequency), provided that the active layer is doped (i.e. $w_0 < d$) and the carrier injection remains negligible. We note that, if the impact of a field-dependent generation rate on the photocurrent is negligible, the same linear $J_{ph}$ versus $w_0$ plot can then also be used to estimate the minority carrier diffusion length from the extrapolated intercept with the $w_0$-axis. If a significant electric field dependence in the generation rate is present, in turn, this is manifested as a strongly voltage, and thus also $w_0$, dependent slope, and a strongly non-linear (super linear) $J_{ph}$ versus $w_0$ dependence is expected. This method is demonstrated experimentally in [69] (see Fig. S3).

## V. CONCLUSIONS

In conclusion, the effect of a doping-induced space-charge region on the charge carrier collection in optically thin solar cells based on low-mobility semiconductors has been clarified. An analytical expression has been derived, explaining the voltage dependence of the photocurrent. Furthermore, the validity of the analytical expression is investigated by numerical simulations based on a drift-diffusion model. Under conditions when the majority carrier conductivity is large enough to screen the electric field inside the neutral region, corresponding to a situation when the charge collection of the minority carriers within the neutral region is dominated by diffusion, the analytical expression shows excellent agreement with numerical simulations and predicts a charge collection efficiency independent of the light intensity. Finally, a good agreement is found between the analytical expression and experimental measurements on organic solar cells based on P3HT:PCBM. Finally, based on our theoretical findings, conditions to avoid doping-induced charge collection losses are discussed and a method to distinguish between field-dependent carrier generation and doping-induced space-charge-limited charge collection is proposed.

**Acknowledgements**

Partial financial support from the Academy of Finland through Project No. 279055, the Jane and Aatos Erkko foundation through project ASPIRE, and Magnus Ehrnrooth Foundation is acknowledged. O.J.S. acknowledges funding from the Swedish Cultural Foundation in Finland and the Sêr Cymru Program through the European Regional Development Fund, Welsh European Funding Office and Swansea University strategic initiative in Sustainable Advanced Materials. S.D. acknowledges funding from the Society of Swedish Literature in Finland. D.S. acknowledges funding from the Magnus Ehrnrooth Foundation. S.W. acknowledges funding through the Research Mobility Programme of Åbo Akademi University.

Supplemental Material for:

# Impact of a doping-induced space-charge region on the collection of photo-generated charge carriers in thin-film solar cells based on low-mobility semiconductors

Oskar J. Sandberg,[1] Staffan Dahlström,[2] Mathias Nyman,[2] Sebastian Wilken,[2,3] Dorothea Scheunemann,[2,3] and Ronald Österbacka[2]

[1]Department of Physics, Swansea University, Singleton Park, Swansea SA2 8PP Wales, United Kingdom

[2]Physics, Center for Functional Materials and Faculty of Science and Engineering, Åbo Akademi University, Porthaninkatu 3, 20500 Turku, Finland

[3]Institute of Physics, Energy and Semiconductor Research Laboratory, Carl von Ossietzky University of Oldenburg, 26111 Oldenburg, Germany

**Experimental**

Organic solar cells with the device structure ITO/MoO$_3$/P3HT:PCBM/LiF/Al were fabricated. Borosilicate glass substrates covered with ITO (Präzisions Glas & Optik GmbH) were used and half of each substrate was etched in aqueous HCl (37-38%) for 40 min. The substrates were cleaned in an ultrasonicator at 60°C in deionized water, acetone and IPA for 10 min each. After cleaning, the substrates were transferred into a nitrogen filled glovebox where the rest of the device fabrication took place. A 2 nm thick layer of MoO$_3$ was thermally evaporated on the ITO substrates. Regioregular (>90 %) poly(3-hexylthiophene-2,5-diyl) (P3HT) and [6,6]-phenyl C61 butyric acid methyl ester (PCBM) from Sigma-Aldrich was dissolved separately in chlorobenzene before mixing at a 1:1 weight ratio. The active layer was spin cast from a 37 mg/ml chlorobenzene solution at 700 rpm for 90 s resulting in a 250 nm thick active layer. The samples were annealed for 10 min at 120°C. The top contact consisting of 0.8 nm LiF and 60 nm Al was thermally evaporated with approximately a 4 mm² overlap with the bottom contact, defining the device area. The active layer thicknesses were measured with atomic force microscopy (AFM).

The charge extraction by a linearly increasing voltage (CELIV) measurements were conducted using a pulse generator (SRS model DG 535) and a function generator (SRS model DS345) for generating the linearly increasing voltage pulse and an oscilloscope (Tektronix TDS 680B) was used to measure the corresponding current response. The sample was mounted in a vacuum cryostat during the measurements. The measurement setup was operated from a computer using a LabVIEW program. The *J-V* measurements were performed in ambient atmosphere using a source-meter (Keithley 2636), an AM1.5 solar simulator (Newport 92250A) was used as light source and neutral density filters with varying optical density were used to reduce the light intensity. The resulting light intensities were measured using a FieldMaster power meter from Coherent.

**Capacitance-voltage measurements**

The measured capacitive extraction current transients are shown in Figure S1. The transients have been measured using the doping-induced charge extraction by linearly increasing voltage (doping-CELIV) method [S1]. In the doping-induced capacitive regime the current transients are determined by the depletion layer (space charge region) capacitance, $j = C_w A = \varepsilon \varepsilon_0 A / w_0$, where $\varepsilon$ is the relative dielectric constant, $\varepsilon_0$ is the vacuum permittivity, $A$ is the slope of the linearly increasing voltage and $w_0$ is the depletion layer width. In the capacitive regime the measured transient currents $j$ are expected to overlap upon varying the applied voltage rise speed $A$, when normalized to the displacement current due to the geometrical capacitance of the device given by $j_0 = \varepsilon \varepsilon_0 A / d$ (where $d$ is the active layer thickness), see Figure S1(a). The normalized current is plotted as a function of the applied voltage and is given by

$$\frac{j}{j_0} = \frac{d}{w_0} = \sqrt{\frac{qN_p d^2}{2\varepsilon\varepsilon_0\left[V_0 - \frac{kT}{q} - V\right]}} \quad (S1)$$

where $q$ is the the elementary charge, $N_p$ is the doping concentration of free holes and the applied transient voltage is given by $V = -At - V_{off}$ for $0 < t < t_{pulse}$, where $V_{off}$ is a steady state offset voltage applied prior to the voltage pulse, $t$ is the time and $t_{pulse}$ is the pulse length. By plotting $\left(\frac{j}{j_0}\right)^{-2}$ as a function of the applied voltage, the doping concentration $N_p$ can be determined from the slope given by $2\varepsilon\varepsilon_0/eN_p d^2$, as depicted by Eq. (S1) [1]. The built-in potential $V_0$ can be directly determined from the intersection with the horizontal axis of the extrapolated line to the linear region, the intersection is given by $-(V_0 - kT/q)$ [1]. We obtain $N_p = 1.4 \times 10^{17}$ cm$^{-3}$ and $V_0 = 0.77$ V.

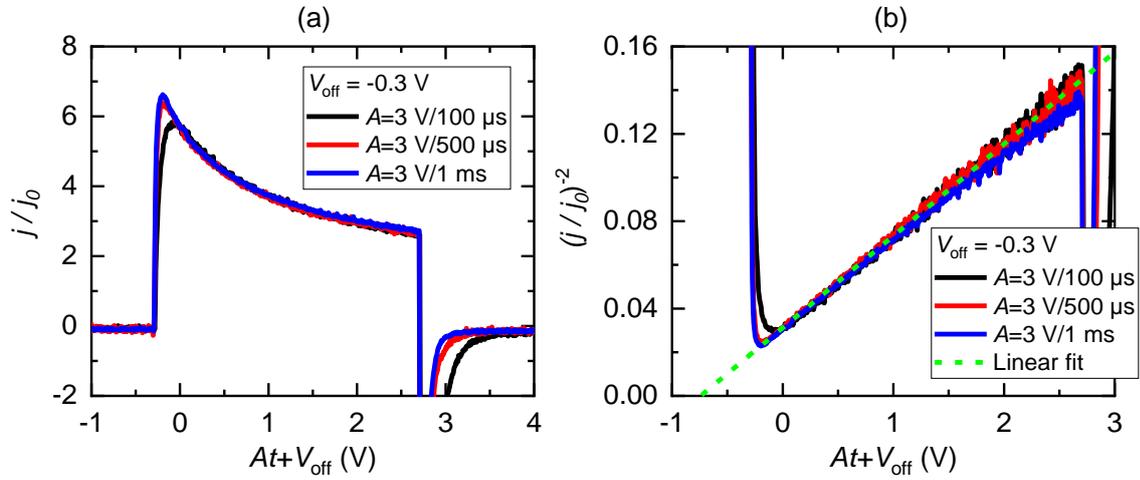

**Figure S1:** Extraction current transients $j$ normalized to $j_0$ is shown in (a). In the doping-induced capacitive regime the transients are overlapping. In (b) $(j/j_0)^{-2}$ is plotted as a function of the applied voltage; $V_0$ is extracted from the intersection of the extrapolated linear fit with the horizontal axis and $N_p$ from the slope of the linear fit.

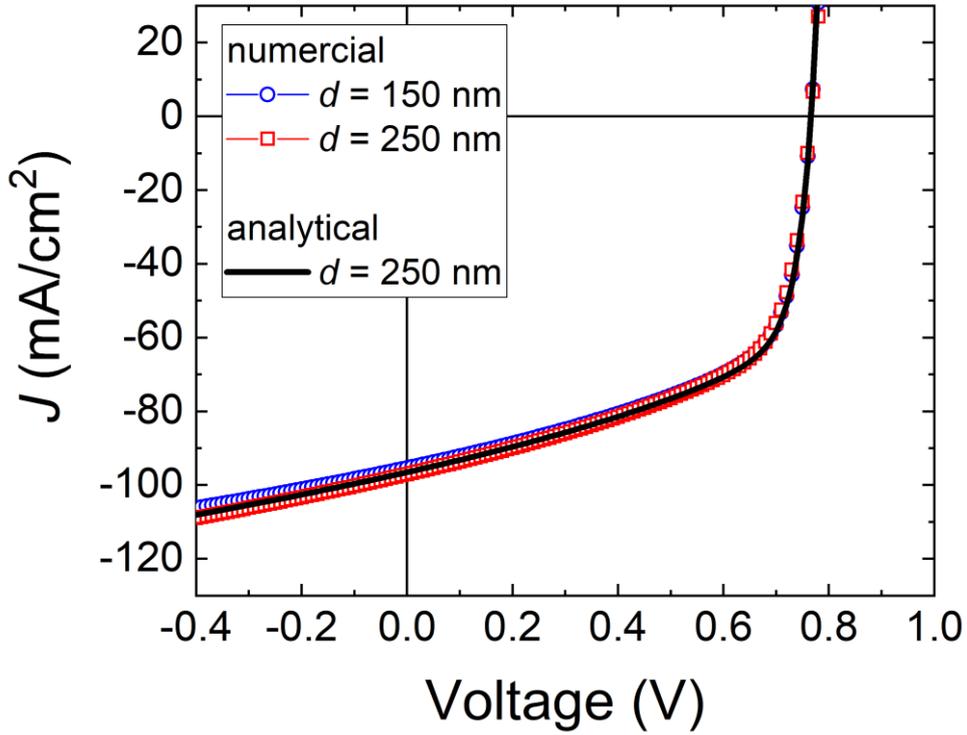

**Figure S2:** Simulated effect of the active layer thickness on the current-voltage characteristics for *p*-doped device under illumination, assuming a uniform generation profile.

**Voltage dependence of the photocurrent**

By plotting the photocurrent as a function of the square root of the applied reverse bias, as seen in Figure S3(a), it is possible to get a strong indication of whether the voltage-dependent photocurrent is caused by doping or other effects, without the need for capacitance-voltage measurements. In the case of doping the photocurrent is linearly dependent on the depletion layer thickness $w_0$ with a corresponding voltage dependence of $\sqrt{V_0 - kT/q - V}$, thereby a linear dependence is seen in Figure S3(a) for large reverse-bias voltages when $|V| \gg V_0 - kT/q$. Provided that this dependence does not change in the limit of low light intensities, this feature is a clear indication that the voltage-dependent photocurrent is caused by doping. Note that for this analysis, only (light-intensity-dependent) *J-V* measurements at high enough reverse bias are required.

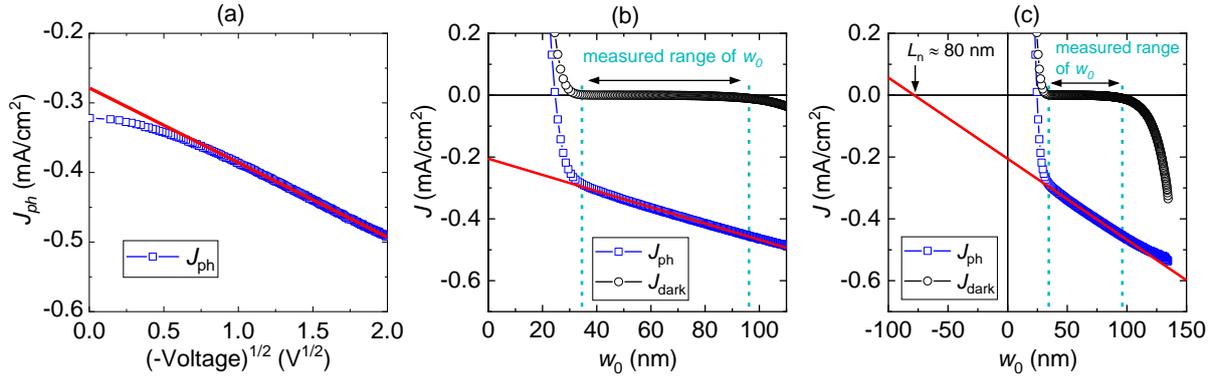

**Figure S3:** In (a) the experimental photocurrent at 0.06 suns (symbols) is plotted as a function of the square root of the applied reverse bias. A linear dependence (red solid line) is seen when the applied voltage in reverse bias is much larger than $V_0 - kT/q$. In (b) and (c): the experimental photocurrent as a function of the depletion layer thickness ($J_{ph}$ versus $w_0$ plot) of the p-doped P3HT:PCBM solar cell (symbols) at 0.06 suns. The linear fit and the corresponding extrapolation in (b) and (c), respectively, are depicted by the red solid lines.

Capacitance-voltage measurements can be conducted in order to confirm the presence of doping in the active layer, as demonstrated in the previous section. The experimental photocurrent as a function of the depletion layer thickness ($J_{ph}$ versus $w_0$ plot) of the *p*-doped P3HT:PCBM solar cell is shown in Figure S3(b) and S3(c). Indeed, a linear dependence (red solid line) is seen within the measured voltage range, suggesting that i) the apparent electric field dependence of the photocurrent originates from the doping-induced space-charge-limited carrier collection, and ii) that the influence of electric field dependent carrier generation rate is negligible. The experimental $w_0$ for outside the measured range is calculated based on the extracted doping concentration and built-in voltage. Note that for $w_0 < 25$ nm, corresponding to $V > 0.5$ V, recombination with injected charge carriers starts to dominate. For $w_0 > 100$ nm, on the other hand, corresponding to $V < -3$ V, the dark current becomes significant compared to the photocurrent and the determined photocurrent is less accurate. In accordance with Eq. (23) in the main text, the minority carrier diffusion length can be estimated from the intercept of the extrapolated the linear fit with the $w_0$-axis, as demonstrated in Figure S3(c).

**Impact of surface recombination of electrons at the anode contact**

The impact of a non-selective contact and surface recombination of electrons at the anode on the short-circuit current density is demonstrated in Figure S4, for different electron mobilities. For the case with non-selective contacts, the electron density at the anode is set equal to $n(0) = n_i^2/N_p$ in the simulations. At high electron mobilities, the losses due to surface recombination are clearly present. In the case when the anode is non-selective for carrier extraction, the contact will act as a sink for both holes and electrons.

This effect can be accounted for in the analytical model by changing the boundary condition at the anode from $J_n(0) = 0$ to $n(0) = 0$ when solving Eq. (12) in the main text; subsequently, the photocurrent is found as

$$J_{ph} = -qG_L\left[w + L_n \tanh\left(\frac{d-w}{2L_n}\right)\right] \tag{S2}$$

for $0 < w < d$. Eq. (S2) shows good agreement with the simulated photocurrent for the device with non-selective contacts. Obviously, in the limit of strong bulk recombination, the photocurrent reduces to $J_{ph} = -qG[w + L_n]$, independent of the surface recombination at the anode. In contrast, when the bulk recombination within the neutral region is weak, the photocurrent approximates $J_{ph} \approx -qG[w + d]/2$ (for $w < d$), which is independent of the diffusion length even though $J_{ph} < qGd$. Concomitantly, the importance of the contacts increases with higher mobilities and lower recombination rates within the bulk, the surface recombination of electrons (at the non-selective anode contact) becoming significant when $L_n > (d-w)/2$. In Figure S4, this corresponds to electron mobilities larger than $\mu_n \approx 10^{-3}$ cm²/Vs. Note that a fixed hole mobility of $\mu_p = 10^{-3}$ cm²/Vs has been assumed in the simulations.

It should be stressed, however, that in case of an Ohmic anode contact, the upwards energy level bending present at this contact generally tends to reduce the surface recombination; for a generic anode contact, the photocurrent thus lies somewhere in between the two extreme cases Eq. (14) from the main text (assuming selective contacts) and Eq. (S2) (assuming non-selective anode contact).

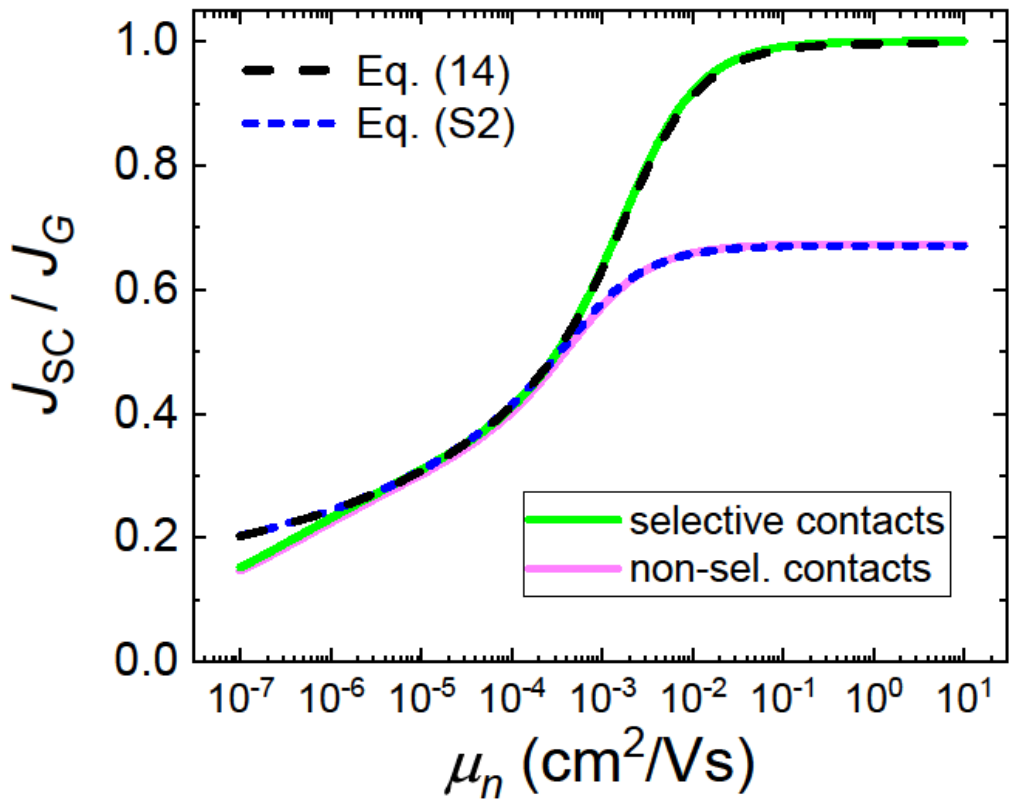

**Figure S4:** The collection efficiency at short-circuit conditions $J_{SC}/J_G$ is simulated as a function of the electron mobility $\mu_n$ at 1 sun light intensity for the case with selective and non-selective contacts, as indicated by the solid lines. A doping concentration of $N_p = 10^{17}$ cm$^{-3}$ is assumed, whereas the recombination coefficient and the hole mobility is fixed at $\beta = 1.2 \times 10^{-11}$ cm³/s and $\mu_p = 10^{-3}$ cm²/Vs, respectively.